\documentstyle[preprint,aps,epsfig]{revtex}
\tightenlines
\begin{document}

\title{How Recent is Cosmic Acceleration?}

\author{\normalsize{Philip D. Mannheim\footnote{Email address: 
mannheim@uconnvm.uconn.edu}} \\
\normalsize{Department of Physics,
University of Connecticut, Storrs, CT 06269} \\
\normalsize{(astro-ph/0104022 v2, June 21, 2001)} \\}

\maketitle

\begin{abstract}
Possibly the most peculiar expectation of the standard fine-tuned 
cosmological paradigm is that cosmic acceleration is to only be a very recent
($z<1$) phenomenon, with the universe being required to be decelerating at
all higher redshifts. Detailed exploration of the Hubble plot out to $z=2$
or so will not only provide an absolute test of this expectation but will
also allow for testing of conformal gravity, a non fine-tuned alternate
cosmological theory which provides equally good fitting to the current $z<1$
Hubble plot data while requiring the universe to be accelerating at higher
$z$ instead. With both standard and conformal gravity being found to be
compatible with a very recently reported $z=1.7$ data point, additional data
points will thus be needed to determine whether the universe is accelerating
or decelerating above $z=1$. 
\end{abstract}

\section{The Hubble plot of standard cosmology}
\medskip

A variety of recent observational measurements
\cite{Riess1998,Perlmutter1998,Bahcall2000,deBernardis2000} have sharply
constrained the space available to the standard model cosmological parameters.
Specifically, phenomenological data fitting using the standard
Einstein-Friedmann cosmological evolution equation	
\begin{equation}
\dot{R}^2(t) +kc^2=\dot{R}^2(t)(\Omega_{M}(t)+\Omega_{\Lambda}(t))
\label{1}
\end{equation}
(where $\Omega_{M}(t)=8\pi G\rho_{M}(t)/3c^2H^2(t)$ is due to ordinary
$\rho_M(t) \sim 1/R^n(t)$ matter and where $\Omega_{\Lambda}(t)=8\pi
G\Lambda/3cH^2(t)$ is due to a cosmological constant $c\Lambda$) is found to
favor current era values $\Omega_{M}(t_0) \simeq 0.3$ and
$\Omega_{\Lambda}(t_0) \simeq 0.7$, with the current era deceleration parameter
$q_0=q(t_0)=(n/2-1)\Omega_{M}(t_0)-\Omega_{\Lambda}(t_0)$ then having to be
of order $-1/2$. While these values are very supportive of the flat 
$\Omega_{k}(t)=-kc^2/\dot{R}^2(t)=1-\Omega_{M}(t)-\Omega_{\Lambda}(t)=0$
inflationary universe paradigm \cite{Guth1981}, they are nonetheless extremely
troubling. Specifically, solving Eq. (\ref{1}) for $\rho_M(t)=B/R^3(t)$ and
$k=0$ yields $R(t)=(B/c\Lambda)^{1/3}sinh^{2/3}(3D^{1/2}t/2)$
where $D=8\pi G\Lambda/3c$, so that
\begin{equation}
\Omega_{\Lambda}(t)=D/H^2(t)=1-\Omega_{M}(t)
=tanh^{2}(3D^{1/2}t/2).
\label{3}
\end{equation}
The identification $c|\Lambda|=\sigma T_V^4$ of the cosmological constant with a
typical particle physics temperature scale $T_V$ of order $10^{16}$ degrees
(viz. $D=4.7 \times 10^{22}$ sec$^{-2}$) would then yield (for $H_0=H(t_0)=65$
km/sec/Mpc) a value for $\Omega_{\Lambda}(t_0)$ of order $10^{59}$, a
value not only overwhelmingly larger than its obtained fitted value, but one not
at all compatible with the bounded $tanh^{2}(3D^{1/2}t_0/2)$ required by Eq.
(\ref{3}). Or, alternatively, if an $\Omega_{M}(t_0)$ of order one is taken as a
given, an associated $\Omega_{\Lambda}(t_0)$ of order $10^{60}$ could then only
be reconciled with Eq. (\ref{1}) in the event that $\Omega_k(t_0)$ was of order
$-10^{60}$ and thus nowhere near flat.     

To get round this problem the standard paradigm then proposes that instead of
using such a particle physics based $D$ one should instead, and despite the
absence of any currently known justification, fine-tune $D$ down by about 60
orders of magnitude and replace it by the phenomenological $D=3.1 \times
10^{-36}$ sec$^{-2}$ with the values $\Omega_{M}(t_0)=0.3$,
$\Omega_{\Lambda}(t_0)=0.7$ then ensuing. However, in its turn, such a proposal
then engenders a further fine-tuning problem for the standard model since for
such a value of $D$ the early universe associated with Eq. (\ref{1}) would need
to be one in which $\Omega_{M}(t=t_{PL})$ would have had to be have been
incredibly close to one at the Planck time $t=t_{PL}$, while
$\Omega_{\Lambda}(t=t_{PL})$ would have had to have been as small as
$O(10^{-120})$. In fact, given such initial conditions, the universe would then
be such that it would decelerate ($q(t)>0$) continually in all epochs until the
cosmological constant finally manages to catch up with the red-shifting matter
density, something which for the phenomenologically chosen value for $D$ would
occur at the incredibly late $z=0.67$ when $q(t)$ would at long last finally
change sign. While it is very peculiar that such a turn around is to occur just
in our particular epoch, nonetheless, independent of one's views regarding the
merits or otherwise of such a proposal, the proposal itself is actually readily
amenable to testing, with a modest increase in the range of $z$ (say to $z=2$)
in the $(d_L,z)$ Hubble plot being able to reveal the presence of any possible
such  turn around. Moreover, such a study would be completely independent of any
dynamical assumptions (such as those required for the (complementary) extraction
of cosmological parameters from the structure of the cosmic microwave background)
and would thus be completely clear cut. Thus in and of itself it would be
extremely informative to extend the range of the Hubble plot. However, as we now
show, it would be of additional interest since it would allow for a rather
unequivocal comparison between standard cosmology and the recently proposed
alternate conformal cosmology, a theory where cosmic acceleration is not at all
of recent vintage.

\section{The Hubble plot of conformal cosmology}

Given the fine-tuning needs of the standard cosmology, it is of value to explore
candidate alternate cosmologies both in and of themselves and also as a
(potentially instructive) foil to the standard theory itself. Of such possible
alternate theories conformal gravity (viz. gravity based on the fully
covariant, locally conformal invariant Weyl action 
$I_W=-\alpha_g \int d^4x (-g)^{1/2} C_{\lambda\mu\nu\kappa} 
C^{\lambda\mu\nu\kappa}$
where $C^{\lambda\mu\nu\kappa}$ is the conformal Weyl tensor and where
$\alpha_g$ is a purely dimensionless gravitational coupling constant) is
immediately suggested since it possesses an explicit symmetry (conformal
invariance) which when unbroken would require the cosmological constant to
vanish. The cosmology associated with the conformal gravity theory was first
presented in \cite{Mannheim1992} where it was shown to possess no flatness
problem, to thus release conformal cosmology from the need for the copious
amounts of cosmological dark matter required of the standard theory.
Subsequently \cite{Mannheim1996,Mannheim1998}, the
cosmology was shown to also possess no horizon problem, no universe age
problem, and, through negative spatial curvature, to naturally lead to cosmic
repulsion.$^{\cite{footnote1}}$ Finally, it was shown
\cite{Mannheim1999,Mannheim2000} that even after the conformal symmetry is
spontaneously broken by a $\Lambda$ inducing cosmological phase transition, the
cosmology is still able to control the contribution of the induced cosmological
constant to cosmic evolution even in the event that $\Lambda$ is in fact as big
as particle physics suggests, to thereby provide a completely natural solution
to the cosmological constant problem. In the present paper we show that this
control actually enables us to provide for a complete and explicit accounting
of the recent high $z$ supernovae Hubble plot data without the need for any
fine tuning at all.   

To explicitly analyze conformal cosmology it is convenient to consider the
generic conformal matter action
\begin{equation}
I_M=-\hbar\int d^4x(-g)^{1/2}[S^\mu S_\mu/2 -
S^2R^\mu_{\phantom{\mu}\mu}/12 + \lambda S^4 +
i\bar{\psi}\gamma^{\mu}(x)(\partial_\mu + \Gamma_\mu(x))\psi -
gS\bar{\psi}\psi] 
\label{5}
\end{equation}
for massless fermions and a conformally coupled order
parameter scalar field. For such an action, when the scalar field acquires a
non-zero expectation value
$S_0$, the entire energy-momentum tensor of the theory is found (for a perfect
matter fluid
$T^{\mu\nu}_{kin}$ of fermions) to take the form
\cite{Mannheim1992,Mannheim1996,Mannheim1998,Mannheim1999,Mannheim2000}        
\begin{equation}
T^{\mu\nu}=T^{\mu\nu}_{kin}-\hbar S_0^2(R^{\mu\nu}-
g^{\mu\nu}R^\alpha_{\phantom{\alpha}\alpha}/2)/6            
-g^{\mu\nu}\hbar\lambda S_0^4,
\label{6}
\end{equation}
with the complete solution to the scalar, fermionic and 
gravitational field equations of motion in a background Robertson-Walker 
geometry (viz. a geometry in which $C^{\lambda\mu\nu\kappa}=0$) then reducing
to  just one relevant equation, viz. $T^{\mu\nu}=0$, a remarkably simple
condition which immediately fixes the zero of energy. We thus see that the
evolution equation of conformal cosmology looks identical to that of standard
gravity save only that the quantity $-\hbar S_0^2 /12$ has replaced the familiar
$c^3/16 \pi G$, so that instead of being attractive the effective cosmological
$G_{eff}=-3c^3/4\pi \hbar S_0^2$ is actually negative to thus naturally lead to
cosmic repulsion, and instead of being fixed as the standard low energy
$G$, the cosmological $G_{eff}$ is instead fixed by the altogether different
scale $S_0$ to thus enable it to be altogether smaller than
$G$.$^{\cite{footnote2}}$ As we shall see below, it is precisely the
replacing of the cosmological $G$ by an altogether smaller $G_{eff}$ that 
enables conformal gravity to naturally solve the cosmological constant problem. 

Given the equation of motion $T^{\mu \nu}=0$, the ensuing conformal cosmology
evolution equation is then found (on setting $\Lambda=\hbar\lambda S^4_0$) to
take a form quite similar to Eq. (\ref{1}), viz. 
\begin{eqnarray}
\dot{R}^2(t) +kc^2
=\dot{R}^2(t)(\bar{\Omega}_{M}(t)+
\bar{\Omega}_{\Lambda}(t)) 
\label{7}
\end{eqnarray}
where $\bar{\Omega}_{M}(t)=8\pi G_{eff}\rho_{M}(t)/3c^2H^2(t)$,
$\bar{\Omega}_{\Lambda}(t)=8\pi G_{eff}\Lambda/3cH^2(t)$. Further, unlike the 
situation in the standard theory where preferred values for the relevant
evolution parameters (such as the magnitude and even the sign of $\Lambda$) are
only determined by the data fitting itself, in conformal gravity essentially
everything is already a priori known. With conformal gravity not needing dark
matter to account for  non-relativistic issues such as galactic rotation curve
systematics \cite{Mannheim1997}, $\rho_{M}(t_0)$ can be determined directly from
luminous matter alone, with galaxy luminosity counts giving a value for it of 
order $0.01\times 3c^2H^2_0/8\pi G$ or so. Further, with $c\Lambda$ being
generated by particle physics vacuum breaking in an otherwise scaleless theory,
since such breaking lowers the energy density, $c\Lambda$ must unambiguously be
negative, with it thus being typically given by $-\sigma T_V^4$ where $T_V$ is
a necessarily particle physics sized scale. Then with $G_{eff}$ also being
negative, the quantity $\bar{\Omega}_{\Lambda}(t)$ must thus be positive, just
as needed to give cosmic acceleration
($q(t)=(n/2-1)\bar{\Omega}_{M}(t)-\bar{\Omega}_{\Lambda}(t)$). Similarly, the
sign of the spatial 3-curvature $k$ is known from theory \cite{Mannheim2000} to
be negative, something which has been independently confirmed from a
phenomenological study of galactic rotation curves \cite{Mannheim1997}.
Moreover, since $G_{eff}$ is negative, the cosmology is singularity free and
thus expands from a (negative curvature supported) finite maximum temperature
$T_{max}$, a temperature which is necessarily greater
\cite{Mannheim1998,Mannheim2000} (and potentially even much greater
\cite{Mannheim1999}) than $T_V$. And finally, with $G_{eff}$ being negative,
the quantity $\bar{\Omega}_M(t)$ must be negative for ordinary $\rho_{M}(t)>0$
matter, with $q(t)$ thus being negative in all epochs.$^{\cite{footnote2a}}$
Consequently in the conformal theory we never need to fine tune in order to
make any particular epoch such as our own be an accelerating
one, and as we shall show below, the conformal theory not
only gives some current era acceleration but in fact gives just the amount
needed to account for the currently available Hubble plot data.    

Given only that $\Lambda$, $k$ and $G_{eff}$ are in fact all negative in the
conformal theory, the evolution of the theory is then completely determined,
with the expansion rate being found
\cite{Mannheim1998,Mannheim1999,Mannheim2000} to be given by
\begin{equation}
R^2= -k(\beta-1)/2\alpha
-k\beta sinh^2 (\alpha^{1/2} ct)/\alpha,~~T_{max}^2/T^2=
1+2\beta sinh^2 (\alpha^{1/2} ct)/(\beta-1),
\label{8}
\end{equation}
where $\alpha c^2=-2\lambda S_0^2=8\pi G_{eff}\Lambda/3c$,  
$\beta=(1- 16A\lambda/k^2\hbar c)^{1/2}=(1+T_V^4/T_{max}^4)/(1-T_V^4/T_{max}^4)$.
In terms of the parameters $T_{max}$ and $T_V$ we thus obtain
\begin{equation}
\bar{\Omega}_{\Lambda}(t)= 
(1-T^2/T_{max}^2)^{-1}(1+T^2T_{max}^2/T_V^4)^{-1},~~ 
\bar{\Omega}_M(t)=-(T^4/T_V^4)\bar{\Omega}_{\Lambda}(t)
\label{9}
\end{equation}
at any $T(t)$ without any approximation at all. From Eq. (\ref{9}) we thus
immediately see that simply because $T_{max}$ is overwhelmingly larger
than the current temperature $T(t_0)$, i.e. simply because the universe is as
old as it is, it automatically follows, without any fine-tuning at all, that the
current era $\bar{\Omega}_{\Lambda}(t_0)$ has to lie somewhere between zero and
one today no matter how big (or small) $T_V$ might actually be, with conformal
gravity thus having total control over the contribution of the cosmological
constant to cosmic evolution. Conformal gravity thus solves the cosmological
constant problem by quenching $\bar{\Omega}_{\Lambda}(t_0)$ rather than by
quenching $\Lambda$ itself (essentially by having a $G_{eff}$ which is
altogether smaller than the standard $G$), and with it being the quantity
$\bar{\Omega}_{\Lambda}(t_0)$ which is the one which is actually measured in
cosmology, it is only its quenching which is actually needed. With conformal
gravity thus being able to naturally accommodate a large $\Lambda$ we are now
actually free to allow $T_V$ to be as large as particle physics suggests. Then,
for such a large $T_V/T(t_0)$ we see that the quantity $\bar{\Omega}_M(t_0)$ has
to be completely negligible today$^{\cite{footnote3}}$ so that $q_0$ must thus,
without any fine-tuning at all, necessarily lie between zero and minus one
today notwithstanding that $T_V$ is huge. Moreover, noting that 
\begin{equation}
tanh^2 (\alpha^{1/2} ct)=(1-T^2/T_{max}^2)/(T_{max}^2T^2/T_V^4+1),
\label{10}
\end{equation}
we immediately see that the current era $\bar{\Omega}_{\Lambda}(t_0)$ is given
by the completely bounded $tanh^2 (\alpha^{1/2} ct_0)$ (so that the
current era curvature contribution to cosmic expansion is then given as 
$\Omega_{k}(t_0)=sech^2 (\alpha^{1/2} ct_0)$), with the current deceleration
parameter being given by the nicely bounded $q_0=-tanh^2 (\alpha^{1/2} ct_0)$.
The essence of the conformal gravity approach then is not to change the matter
and energy content of the universe but rather only their effect on cosmic
evolution, with the cosmological constant itself no longer needing to be
quenched.

While completely foreign to standard gravity, a universe in which
$\rho_M(t)$ makes a completely negligible contribution to current era
cosmic evolution is, as we now show, nonetheless fully compatible with the
currently available $z<1$ Hubble plot data. Specifically, through use of type Ia
supernovae the authors of \cite{Riess1998,Perlmutter1998} were able to measure
the dependence of luminosity distance $d_L$ on redshift out to $z=1$. To fit
their data we thus need to determine the dependence of $d_L$ on $z$ in the
conformal theory, something we can readily do now that we have obtained  the
expansion factor $R(t)$. Thus, in the conformal theory we find first that the
Hubble parameter is given as
\begin{equation}
H(t)=\alpha^{1/2}c(1-T^2(t)/T^2_{max})/tanh(\alpha^{1/2}ct)
\label{11}
\end{equation}
with its current ($T_{max} \gg T(t_0)$) value being found to obey
$-q_0=tanh^2(\alpha^{1/2}ct_0) =\alpha c^2/H^2_0$, with the current age of the
universe then being given by $t_0=arctanh [(-q_0)^{1/2}]/(-q_0)^{1/2}H_0$, viz.
by $t_0 \geq 1/H_0$. For temperatures well below $T_{max}$ and for the
naturally achievable \cite{Mannheim1999} $T_V
\ll T_{max}$ case of most practical interest  to conformal
gravity (viz. a case where
$T_{max}^2T^2(t_0)/T_V^4$ can be of order one) we may set $R(t)= 
(-k/\alpha)^{1/2} sinh(\alpha^{1/2} ct)$, so that for geodesics
$\int_{t_1}^{t_0}c dt/R(t)=\int_0^{r_1}dr /[1-kr^2]^{1/2}$ we obtain 
\begin{equation}
(-k)^{1/2}r_1
=coth(\alpha^{1/2}ct_0)/sinh(\alpha^{1/2}ct_1)-coth(\alpha^{1/2}ct_1)/
sinh(\alpha^{1/2}ct_0).
\label{12} 
\end{equation}
Then, with
$sinh(\alpha^{1/2}ct_1)=(-q_0)^{1/2}/(1+q_0)^{1/2}(1+z)$ where
$z=R(t_0)/R(t_1)-1$, we find that we can express the general luminosity distance
$d_L=r_1R(t_0)(1+z)$ entirely in terms of the current era $H_0$ and $q_0$
according to the very compact relation
\begin{equation}
H_0 d_L/c=-(1+z)^2\left\{1-[1+q_0-q_0/(1+z)^2]^{1/2}\right\}/q_0
\label{13}
\end{equation}
Conformal gravity fits to the luminosity distance can thus be parametrized via
the one parameter $q_0$, a parameter which must lie somewhere between zero and
minus one, with $d_L$ thus having to lie somewhere between
$d_L(q_0=0)=cH_0^{-1}(z+z^2/2)$ and $d_L(q_0=-1)=cH_0^{-1}(z+z^2)$ at 
temperatures well below $T_{max}$. Given Eq. (\ref{13}) we turn now to a data
analysis. 

\section{Fits to the Hubble plot}

For the fitting we shall follow \cite{Perlmutter1998} and fit 38 of their
42 data points together with 16 of the 18 earlier lower $z$ points of
\cite{Hamuy1996},  for a total of 54 data points with reported effective blue
apparent magnitude $m_i$ and uncertainty $\sigma_i$. (While we thus leave out 6
questionable data points for the fitting, nonetheless, for completeness we 
still include them in the figures.) For the fitting we calculate the apparent
magnitude $m$ of each supernova at redshift $z$ via $m=25+M+5log_{10} d_L$
($d_L$ in Mpc) where $M$ is their assumed common absolute magnitude, and
minimize $\chi^2(q_0,M)=\sum (m-m_i)^2/\sigma_i^2$ as a function of the two
parameters $q_0$ and $M$. In all of our fits $M$ (as determined using $H_0=65$
km/sec/Mpc) is found to be in agreement with the analyses of
\cite{Riess1998,Perlmutter1998}, with our best fit $\chi^2=58.62$ being
obtained for $q_0=-0.37$, $M=-19.37$. We display this fit as the upper curve in
Fig. (1) where we also present as the lower curve the corresponding
$\Omega_{M}(t_0)=0.3$, $\Omega_{\Lambda}(t_0)=0.7$ standard model
fit, a fit which gives $\chi^2=57.74$ (and $M=-19.37$) for the same 54 points.
As we thus see, in the detected region the best fits of the two models are
completely indistinguishable, only in fact departing from each other at the
highest available redshifts. Moreover, the conformal cosmology fits turn out to
be extremely insensitive to the actual value of $q_0$, with other typical $q_0,M$
fits being $\chi^2(0,-19.29) =61.49$, $\chi^2(-0.25,-19.34)=58.96$,
$\chi^2(-0.5,-19.40)=59.11$, $\chi^2(-0.75,-19.46)=63.59$, and
$\chi^2(-1.00,-19.54)=75.79$.$^{\cite{footnote3a}}$ The data will thus support
conformal cosmologies with a fairly broad range of negative values of $q_0$
beginning at $q_0=0$. In fact such high quality $q_0=0$ fits (viz. fits with
$R(t)= (-k)^{1/2}ct$) have already been reported earlier,$^{\cite{footnote6}}$
with the authors of \cite{Perlmutter1998} having noted in passing that such
fits were actually as good as their best standard model fits, and with the 
authors of \cite{Dev2000} having presented such
$q_0=0$ fits in an exploration of generic power law expansion rate cosmologies.
What is new here is that we derive such $q_0=0$ fits within the framework of a
well-defined cosmological model while also extending them to non-zero $q_0$
(i.e. to non-zero $\Lambda$). 

As we just noted, current Hubble plot data do not allow us to resolve between
standard and alternate gravity theories. However, since the standard theory is a
decelerating one above $z=1$ while the conformal theory continues to accelerate,
continuation of the Hubble plot beyond $z=1$ will actually enable us to
discriminate between the various options. Thus in Fig. (2) we plot the higher
$z$ expectations. The highest curve in the figure is the conformal gravity fit
for $q_0=-0.37$, the middle curve in the figure is the conformal gravity fit for
$q_0=0$, and the lowest curve in the figure is the $\Omega_{M}(t_0)=0.3$,
$\Omega_{\Lambda}(t_0)=0.7$ standard model expectation. As we see, the curves
start to depart from each other fairly rapidly once we go above $z=1$, with the
three cases respectively yielding $m=27.17$, $m=27.04$ and $m=26.75$ at $z=2$, a
difference of at least $0.3$ magnitudes between standard gravity and the
conformal alternative. (At $z=5$ the respective magnitudes are $m=30.40$,
$m=30.25$ and $m=29.14$.) Since conformal gravity handles the cosmological
constant problem, the primary problem troubling the standard theory, so readily 
(its other successes and its own difficulties are discussed in
\cite{Mannheim1998,Mannheim1999}) it would thus appear to merit further
consideration, with only a modest extension of the Hubble plot readily
enabling us to discriminate between standard gravity and its conformal
alternative while potentially even being definitive for 
both.$^{\cite{footnote7}}$ The author is indebted to Drs. G. V. Dunne, K.
Horne, D. Lohiya, R. Plaga and B. E. Schaefer for helpful comments. This work
has been supported in  part by the Department of Energy under grant No.
DE-FG02-92ER40716.00.

\section {Added Note} 

Since this paper was first released it was announced \cite{Riess2001}
that new analysis of SN 1997ff now puts this supernova at
$z=1.7^{+0.1}_{-0.15}$ and establishes a lower bound on its luminosity large
enough to definitively exclude the two most commonly considered
non-cosmological explanations (dust extinction and intrinsic luminosity
evolution) of the supernovae Hubble plot data. Interestingly, this lower bound
turns out to be very close not only to the values ($m=26.52$, $m=26.64$)
expected of $q_0=0$ and $q_0=-0.37$ conformal gravity at $z=1.7$ but also to
the nearby $m=26.32$ expectation of an
$\Omega_M(t_0)=0.3$, $\Omega_{\Lambda}(t_0)=0.7$ standard model at that $z$,
thus making further exploration of the $z>1$ region potentially very
significant. In their analysis the authors of \cite{Riess2001} present their SN
1997ff data in the form of confidence contours in distance modulus ($m-M$)
versus redshift space. To illustrate their data we have augmented Fig. (2) by
adding in the 68\% and 95\% confidence region values for the apparent magnitude
$m$ (as obtained from the quoted $m-M$ by adding $M=-19.37$) at redshifts
$z=1.65$, $z=1.7$ and $z=1.75$, and have presented an enlarged view of the
relevant region in Fig. (3). (In these figures the two inner horizontal bars on
the vertical data points represent the extent of the 68\% confidence region at
each of the chosen redshifts while the two outer bars represent the 95\%
confidence one.) To estimate the effect of SN 1997ff on the
$\hat{\chi}^2(q_0,M)$ that we obtained for our fits to the $z<1$ data we have
calculated the change in $\Delta \hat{\chi}^2(q_0,M)$ generated by an effective
additional data point with the 68\% confidence region value of $m=25.83 \pm
0.5$ at redshift $z=1.7^{+0.1}_{-0.15}$, to find typical
values $\Delta \hat{\chi}^2(0,-19.28)=1.60$ and $\Delta
\hat{\chi}^2(-0.33,-19.35)=2.08$ to be compared with
$\Delta\hat{\chi}^2(-19.36)=0.83$ for the standard model. 

While the conformal gravity $\Delta\hat{\chi}^2$ values are actually less than
those generated by some of the outlier points in the $z<1$
data,$^{\cite{footnote8}}$ it is important to note that as well as possessing
statistical errors the SN 1997ff data happen to also possess an explicitly
identified systematic error. Specifically SN 1997ff just happens to be lensed
by two foreground galaxies at $z=0.56$ both of which lie very close to the line
of sight, with the authors of \cite{Riess2001} estimating that this would cause
SN 1997ff to appear 0.2 magnitudes or so brighter than it actually is. Dimming
the quoted distance modulus data by this amount would serve to move conformal
gravity into the 68\% confidence region with the modified $\Delta
\hat{\chi}^2(q_0,M)$ then becoming $\Delta \hat{\chi}^2(0,-19.28)=0.82$ and
$\Delta\hat{\chi}^2(-0.33,-19.35)=1.16$ to be compared with a modified
$\Delta\hat{\chi}^2(-19.36)=0.29$ for the standard model. Thus we see that at
present it is not yet possible to discriminate between standard gravity and the
conformal alternative, with both of the theories being compatible with the SN
1997ff data. In fact, given that there is a systematic lensing effect (and as
yet only one $z>1$ data point of course), we see that since the dimmed data
straddle the coasting ($q(t)=0$ at all times) universe (equivalent to $q_0=0$
conformal gravity), at the present time one is not in fact able to determine
whether the universe is actually accelerating or decelerating above
$z=1$,$^{\cite{footnote9}}$ with further data (such as those anticipated
in the upcoming SNAP satellite mission) thus being needed to resolve the issue
and identify any specific trend in the $z>1$ region that might exist.

\begin{figure} 
\epsfig{file=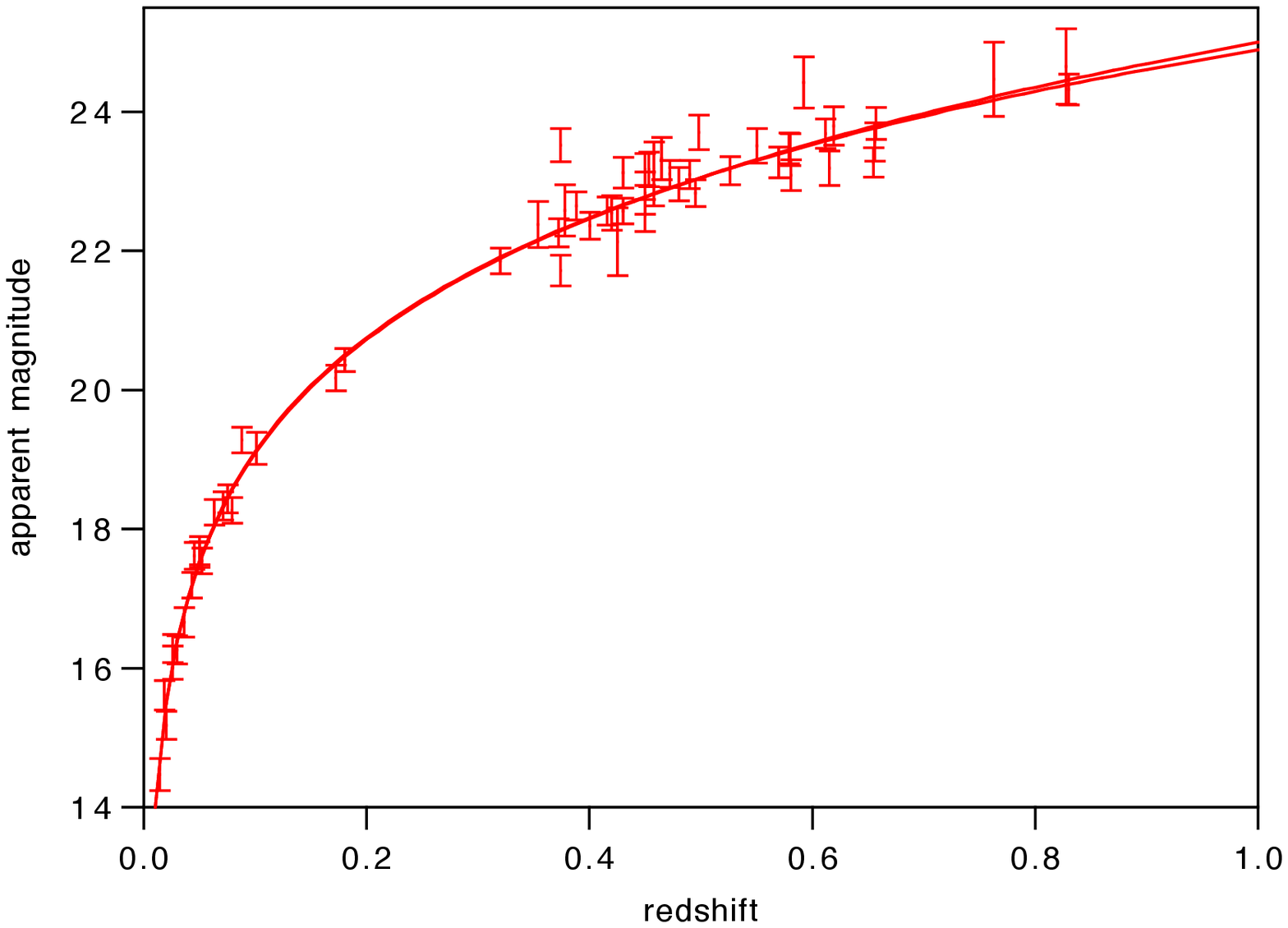}
\caption{The $q_0=-0.37$ conformal gravity fit (upper curve) and the  
$\Omega_{M}(t_0)=0.3$, $\Omega_{\Lambda}(t_0)=0.7$ standard model fit (lower
curve) to the $z<1$ supernovae Hubble plot data.}
\label{f1}
\end{figure}

\begin{figure}
\epsfig{file=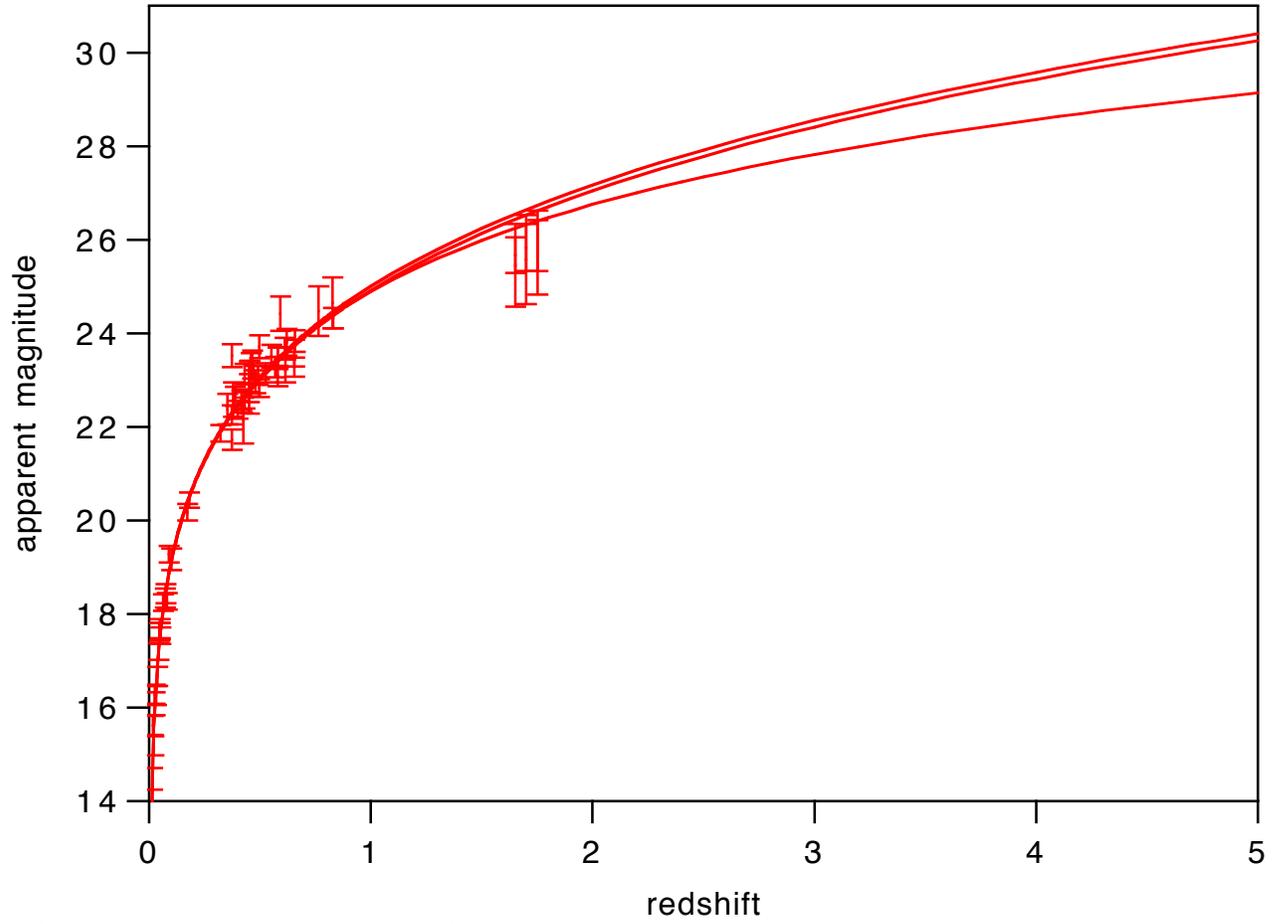}
\caption{Hubble plot expectations for $q_0=-0.37$ (highest curve) and
$q_0=0$ (middle curve) conformal gravity and for
$\Omega_{M}(t_0)=0.3$,
$\Omega_{\Lambda}(t_0)=0.7$ standard gravity (lowest curve).}
\label{f2}
\end{figure}

\begin{figure}
\epsfig{file=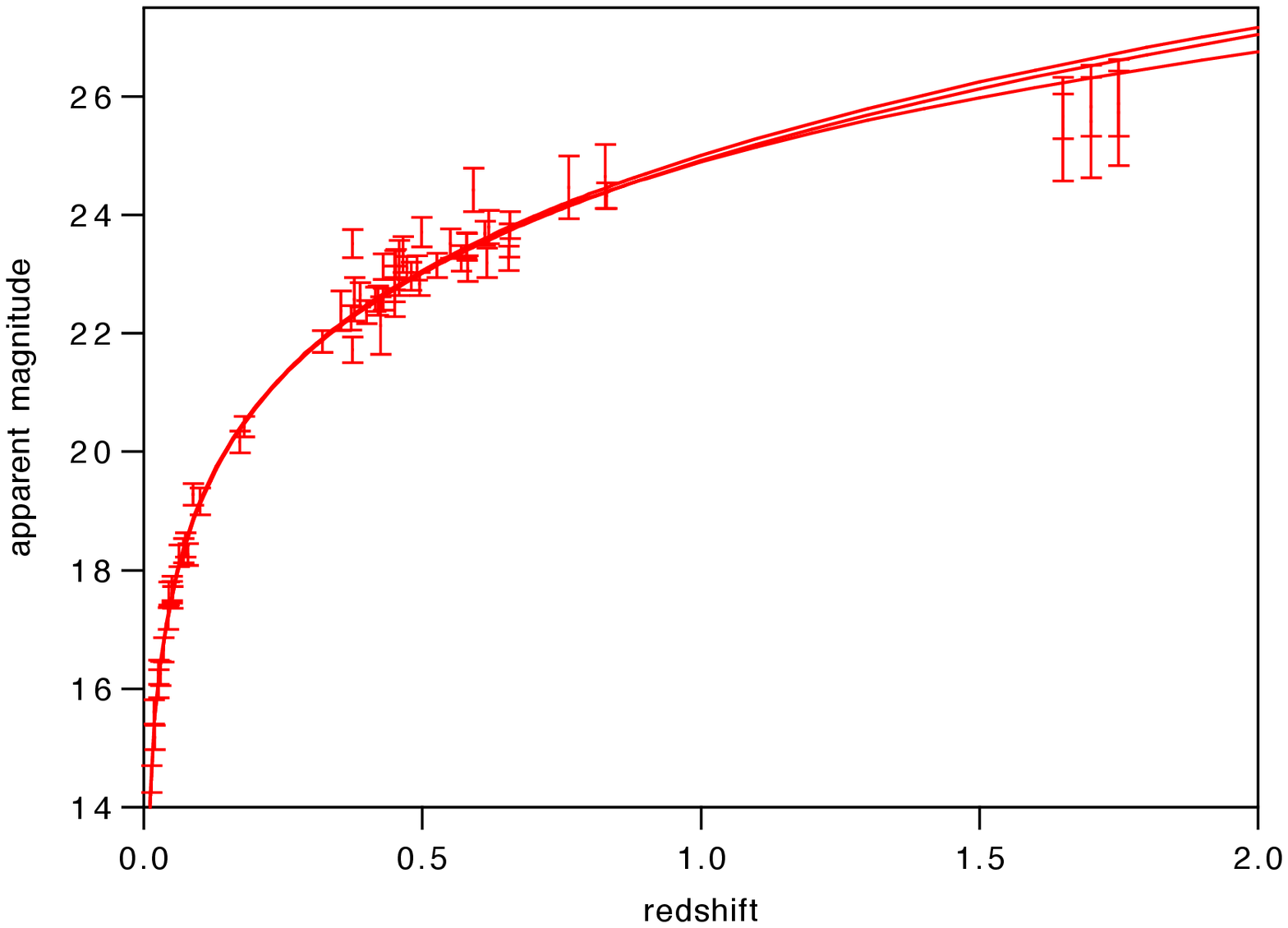}
\caption{The $q_0=-0.37$ (highest curve) and $q_0=0$ (middle curve) conformal
gravity fits and the $\Omega_{M}(t_0)=0.3$, $\Omega_{\Lambda}(t_0)=0.7$
standard model fit (lowest curve) to the $z<2$ Hubble plot data.}
\label{f3}
\end{figure}


\begin{thebibliography}{99}
\bibitem{footnote2a} Included in this class of $q(t)<0$ universes are the  
coasting ones in which $q(t)=0^{-}$.
\bibitem{footnote3}$\bar{\Omega}_M(t_0)$ is suppressed by 
$G_{eff}$ being small, and not by $\rho_{M}(t_0)$ itself being small, with
$G_{eff}$ being made smaller the larger rather than the smaller 
$S_0$ gets to be, to thus enable the $c\Lambda/\rho_{M}(t_0) =
\bar{\Omega}_{\Lambda}(t_0)/\bar{\Omega}_M(t_0)$ ratio to be as large as
particle physics suggests while not leading to any 60 order of magnitude
conflict with observation.
\bibitem{footnote3a} Since redshift position uncertainties $\Delta
z_i$ are also reported in the supernovae data, for completeness we have also
minimized $\hat{\chi}^2(q_0,M)=\sum (m-m_i)^2/(\sigma_i^2+\sigma_{z_i}^2)$ where
$\sigma_{z_i}^2=[5log_{10} d_L(q_0,z_i+\Delta z_i)-5log_{10}
d_L(q_0,z_i)]^2/2+[5log_{10} d_L(q_0,z_i-\Delta z_i)-5log_{10}
d_L(q_0,z_i)]^2/2$ is the effect of the redshift uncertainty, with it being the
fitting procedure itself which is used to normalize $\sigma_{z_i}$ to the
apparent magnitude. With this prescription we then find that our best conformal
gravity fit is now given as $\hat{\chi}^2(-0.33,-19.35)=54.13$ for the 54
points, to be compared with the corresponding best\ $\Omega_{M}(t_0)=0.3$,
$\Omega_{\Lambda}(t_0)=0.7$ standard model fit of $\hat{\chi}^2(-19.36)=53.27$.
Moreover, the conformal cosmology fits continue to be insensitive to the actual
value of $q_0$, with other typical $q_0,M$ fits then being
$\hat{\chi}^2(0,-19.28) =56.00$,
$\hat{\chi}^2(-0.25,-19.33)=54.26$, $\hat{\chi}^2(-0.5,-19.39)=54.78$,
$\hat{\chi}^2(-0.75,-19.47)=59.01$, and $\hat{\chi}^2(-1.00,-19.55)=69.88$.
\bibitem{footnote6}While such $R(t)=(-k)^{1/2}ct$ fits would have to be
associated with an $\Omega_M(t_0)=0$, $\Omega_{\Lambda}(t_0)=0$ empty universe
in the standard model, in the conformal theory they would instead be associated
with a current era $\bar{\Omega}_M(t_0)=0$, $\bar{\Omega}_{\Lambda}(t_0)=0$
universe, viz. one with its normal fill of matter but  which is now in a
curvature dominated phase.
\bibitem{footnote7}With $k$ explicitly being negative in conformal
gravity, determining its predictions for the temperature dependence of the
cosmic microwave background is thus also paramount for the theory.
\bibitem{footnote8} Even in the standard model fits to the data below $z=1$ 
some 10 of the 54 fitted points are each individually found to contribute an
amount in excess of 2 units to the total $\hat{\chi}^2(-19.36)=53.27$ (the
largest single individual contribution goes as high as $7.10$), with all of
these particular points thus lying quite a bit away from the expected standard
model value. With there thus being a lot of scatter in the Hubble plot data one
can only extract limited information from any given single data point.
\bibitem{footnote9} In fact it is only the assumption of the a priori validity
of standard gravity which leads to the favoring of deceleration above $z=1$.
\end{thebibliography}
\end{document}